\documentstyle[12pt,epsfig]{article}
\begin{document}
\baselineskip = 6.0mm
\topmargin= -5mm
 \begin{center}
\begin{large}
 {\bf { No massless boson 

in chiral symmetry breaking in NJL and Thirring models }}

\end{large}

\vspace{2cm}

   Makoto HIRAMOTO  and  Takehisa FUJITA

Department of Physics, Faculty of Science and Technology  
  
Nihon University, Tokyo, Japan

\vspace{3cm}

{\large ABSTRACT} 
 
\end{center}

We show that the chiral symmetry breaking occurs in the vacuum of the massless 
Nambu-Jona-Lasinio (NJL) and Thirring  models without a Goldstone boson. 
The basic reason of non-existence of the massless boson is due to the 
fact that the new vacuum after the symmetry breaking acquires nonzero fermion mass 
which inevitably leads to massive bosons. The new vacuum has a finite condensate 
of $ \langle \bar \psi \psi \rangle $ with 
the chiral current conservation.  Thus, it 
contradicts the Goldstone theorem, and we show that the proof of the Goldstone 
theorem cannot be justified any more for fermion field theory models 
with regularizations.

\vspace{1cm}
\noindent
PACS numbers: 11.10.Kk, 03.70.+k, 11.30.-j, 11.30.Rd

\newpage

\begin{enumerate}
\item{\large Introduction}

Symmetries play a most important role for the 
understanding of the basic behavior in quantum field theory. The breaking of 
symmetries is also important and interesting since the vacuum 
can violate the symmetry which is possessed in the Lagrangian. 

In field theory, there is an interesting theorem for the spontaneous symmetry 
breaking. That is, the Goldstone theorem \cite{q1,q2}, and 
it states that there appears 
a massless boson if the symmetry is spontaneously broken. 

For the spontaneous symmetry breaking in fermion field theory, Nambu and 
Jona-Lasinio first pointed out that the current current interaction model 
( NJL model) presents a good example of the spontaneous symmetry breaking \cite{q6}. 
Indeed, they showed in their classic paper that the chiral symmetry is broken 
if they start from the massless fermion Lagrangian which possesses a chiral 
symmetry. Then, they construct the new vacuum 
with the Bogoliubov transformation, and proved that the new vacuum 
breaks the chiral symmetry, and besides they found that the $originally$ 
massless fermion acquires an induced mass. To be more important, 
all of the physical observables like the boson mass is measured 
by the new fermion mass. Up to this point, it was indeed 
quite an interesting and convincing scenario that happened to  
the chiral symmetry breaking phenomena. 

However, Nambu and Jona-Lasinio claimed that there appears 
a massless boson after the chiral symmetry breaking. 
This massless boson, if at all exists, must be constructed 
by the fermion and antifermion by its dynamics. But all of the 
calculations which have been carried out up to now show 
that the boson in the NJL model with the massive fermion 
is always massive, and the boson mass is proportional to the fermion mass. 
Therefore, a massless boson can be obtained only 
when one sets the fermion mass to zero \cite{q07}. But this is not 
allowed since the induced fermion mass can never become zero. 
A massless boson might be obtained when the interaction 
strength is at the strongest limit since 
all the fermion mass could be eaten up by the interaction. 
However, Nambu and Jona-Lasinio claim that there exists 
a massless boson regardless the strength of the coupling constant 
even if the strength of the coupling constant 
is quite small. But obviously, this is physically not acceptable. 

In this paper, we show by explicit calculations that the NJL model has no massless 
boson, and the boson mass is always finite. But the symmetry breaking 
phenomenon is exactly the same as the one presented by Nambu and Jona-Lasinio. 
The new vacuum breaks the chiral symmetry since the energy of 
the new vacuum is lower than the trivial one. 
In this case, the $originally$ massless fermion 
can indeed acquire an induced mass which is measured by the cutoff 
momentum $\Lambda$. With this finite fermion mass, we calculate the boson 
mass, and obtain the mass which is indeed finite, and  
its magnitude certainly depends on the strength of the coupling constant. 
It is interesting to observe that there is no bound state 
of the bosonic state if the strength of the coupling constant is weaker 
than the critical value. 

Here, we also carry out the calculation of the chiral symmetry 
breaking in the massless Thirring model which is a two dimensional 
field theory \cite{q7}. In fact, we show that the new vacuum breaks the chiral 
symmetry, and the massless fermion acquires an induced mass. 
Therefore, the massless Thirring model becomes just the massive 
Thirring model with the induced fermion mass, and it is 
clear that there exists always a massive boson even 
for the weak coupling region \cite{q9}. 
Therefore, even in two dimensional field theory models with the regularization, 
the continuous symmetry is broken, but there appears no massless boson. 
In this respect, the situation in the Thirring model is just the same 
as the NJL model. Since there is no massless boson in the symmetry 
breaking, this is not inconsistent with Coleman's theorem \cite{q3}. 
But we should stress that the symmetry is, in fact, broken in the two 
dimensional field theory models. Further, a recent work \cite{q34} shows  
that a chiral symmetry in the two dimensional QCD with masselss fermions 
is broken without the anomaly, but there appears no massless boson, and 
instaed, there exists a massive boson. 

This means that there must be some problems in the Goldstone theorem 
for the fermion field theory models with the regularization. 
Here, we show that the procedure of proving Goldstone theorem 
for the fermion field theory models cannot be justified any more
when the fermion current is regularized with the point splitting. 
The defect of the Goldstone theorem for the fermion field theory 
with the regularization is essentially based on the fact that 
the boson is a complex object, and it must be constructed by the fermion 
and antifermion. But the fermion current must be regularized, and,  
in this case, the boson cannot be treated as a simple elementary particle. 
Therefore, the method of proving the Goldstone theorem which is successful 
for the boson field theory cannot be applied any more to the fermion field theory 
model with the regularization.

This paper is organized as follows. In the next section, we first treat the Bogoliubov 
transformation in the Nambu-Jona-Lasinio(NJL) model, and obtain a new vacuum. 
The mass of the boson is evaluated in the regularized 
NJL model. In section 3,  we discuss the chiral symmetry breaking 
in the massless Thirring model 
and calculate the boson mass with the Bogoliubov transformed vacuum. 
In section 4, we explain the breaking down of the Goldstone 
theorem for the fermion field theory models with the current regularization. 
Finally, section 5 summarizes 
what we have clarified in this paper.

\vspace{1cm}

\item{\large Nambu-Jona-Lasinio model }

Here, we discuss the four dimensional current-current interaction 
model by Nambu and Jona-Lasinio \cite{q6}. This paper is the original one 
that initiated the chiral symmetry breaking phenomena.  
Here, we treat the NJL model just in the same manner as the one given by 
Nambu and Jona-Lasinio as far as the symmetry breaking mechanism 
is concerned. However, the boson mass determination is different, and 
we carry out the calculation in terms of the Fock space expansion. 
Also, we carry out the RPA calculation, but it turns out that 
the RPA calculation gives the boson spectrum which is similar to 
that calculated by the Fock space expansion. But the boson mass is slightly 
lower than the result of the Fock space expansion as the function 
of the coupling constant.

Now, we carry out the calculation which is based on the Bogoliubov 
transformation, and show that the chiral symmetry is indeed broken.  
However, we also show that there appears no massless boson in this 
regularized NJL model.

Here, we first quantize the fermion field in a box $L^3$ 
$$ \psi({\bf r}) = \frac{1}{\sqrt{L^3}}\sum_{{\bf n},s}
\left[a({\bf n},s)u({\bf n},s)e^{i{2\pi\over L}{\bf n}\cdot{\bf r}}+
b^{\dagger}({\bf n},s)v({\bf n},s)e^{-i{2\pi\over L}{\bf n}\cdot{\bf r}}\right] 
\eqno{ (2.1)} $$
where $s$ denotes the spin index, and $s= \pm 1$. Also, the spinors are defined as 
$$ u({\bf n},s)=\frac{1}{\sqrt{2}}
\left( \begin{array}{c}
{{\mbox{\boldmath $\sigma$}} \cdot \bf{\hat n}}\chi^{(s)} \\
       \chi^{(s)} \end{array}
\right), \quad $$
$$ v({\bf n},s)=\frac{1}{\sqrt{2}}
\left(  \begin{array}{c}
\chi^{(s)} \\
{{\mbox{\boldmath $\sigma$}} \cdot \bf{\hat n}}\chi^{(s)} \end{array}
\right), \quad $$
Now, we define new fermion operators by the Bogoliubov transformation, 
$$ c({\bf n},s) = e^{- {\cal A} } a({\bf n},s) e^{{\cal A}} = 
\cos \left({\theta_{\bf n}\over 2}-{\pi\over 4}\right) a({\bf n},s) 
- s \sin \left({\theta_{\bf n}\over 2}-{\pi\over 4}\right) 
b^{\dagger}(-{\bf n},s) \eqno{(2.2a)} $$
$$ d^{\dagger}(-{\bf n},s) = e^{- {\cal A} } b^{\dagger}(-{\bf n},s) e^{{\cal A}} 
= \cos \left({\theta_{\bf n}\over 2}-{\pi\over 4}\right) b^{\dagger}(-{\bf n},s) + 
s \sin \left({\theta_{\bf n}\over 2}-{\pi\over 4}\right) a({\bf n},s) \eqno{(2.2b)} $$
where the generator of the Bogoliubov transformation is given by
$$ {\cal A} = - \sum_{{\bf n},s} \left({\theta_{\bf n}\over 2}-{\pi\over 4}\right)
 \left( a^{\dagger}({\bf n},s)b^{\dagger}(-{\bf n},s)  -b(-{\bf n},s)
a({\bf n},s) \right)  . \eqno{(2.3)} $$
$ \theta_{\bf n}$ denotes the Bogoliubov angle 
which can be determined by the condition that the vacuum energy is minimized. 
In this case, the new vacuum state is obtained as 
$$ \mid \Omega \rangle = e^{ - {\cal A} }  |0\rangle  .  \eqno{(2.4)}  $$
In what follows, we treat the NJL Hamiltonian with the Bogoliubov transformed 
vacuum state. In order to clearly see some important difference between 
the massive fermion and massless fermion cases, we treat the two cases 
separetely. 

\vspace{1cm}
\begin{enumerate}
\item{Massive fermion case}

The Lagrangian density for the NJL model with the massive fermion can be written as 
$$ {\cal L}= i \bar \psi  \gamma_{\mu} \partial^{\mu}  \psi  -m_0\bar{\psi}\psi 
+G \left[ (\bar{\psi}\psi )^2 
+(\bar{\psi}i\gamma_5\psi )^2  \right]  . \eqno{(2.5)}  $$
Now, we can obtain the new Hamiltonian under the Bogoliubov transformation, 
$$ H= \sum_{{\bf n},s}\left\{|{\bf p}_{\bf n}| \sin{\theta_{\bf n}} 
+\left(m_{0}+\frac{2G}{L^3}\mathcal{B}\right)
\cos{\theta_{\bf n}}\right\}
\left( c^{\dagger}({\bf n},s)c({\bf n},s)+d^{\dagger}(-{\bf n},s)d(-{\bf n},s) 
\right) $$
$$ + \sum_{{\bf n},s}
\left\{-|{\bf p}_{\bf n}|s \cos{\theta_{\bf n}}+
\left(m_{0}+\frac{2G}{L^3}\mathcal{B}\right)s \sin{\theta_{\bf n}} \right\}
\left(c^{\dagger}({\bf n},s)d^{\dagger}(-{\bf n},s)+
d(-{\bf n},s)c({\bf n},s) \right) $$
$$ + {H'}_{int}  \eqno{ (2.6)} $$
where $ {H'}_{int}$ is the interaction term. Since the ${H'}_{int}$ 
is quite complicated, and 
besides its explicit expression is not needed in this context, we will not 
write it here.   
${\mathcal{B}} $ is defined as 
$$ {\mathcal{B}}=\sum_{{\bf n},s}\cos\theta_{\bf n} . $$ 
Now, we can define the renormalized fermion mass $m$
$$ m = m_{0}+\frac{2G}{L^3}\mathcal{B}. \eqno{ (2.7)} $$
The Bogoliubov angle $\theta_{\bf n}$ can be determined by imposing the condition 
that the $cd$ term in eq.(2.6) must vanish. Therefore, we obtain
$$ \tan \theta_{\bf n}  = {|{\bf p}_{\bf n}|\over m} . \eqno{ (2.8)} $$
This Bogoliubov angle $\theta_{\bf n}$ does not change when the mass 
varies from $m_0$ to $m$. 
In this case, the vacuum is just the same as the trivial vacuum of the massive case, 
except that the fermion mass is replaced by the renormalized 
mass $m$. The rest of the theory 
becomes identical to the massive NJL model with the same  
interaction Hamiltonian ${H'}_{int}$. Therefore, there is no symmetry breaking, 
and this vacuum has no condensate. 

\item{Massless fermion case}

Here, we present the same procedure for the massless fermion 
case in order to understand why the fermion has to become massive. 

We start from the Lagrangian density with no mass term in eq.(2.5). 
Under the Bogoliubov transformation, we obtain the new Hamiltonian 
$$ H= \sum_{{\bf n},s}\left\{|{\bf p}_{\bf n}| \sin{\theta_{\bf n}} 
+\frac{2G}{L^3}\mathcal{B} 
\cos{\theta_{\bf n}}\right\}
\left( c^{\dagger}({\bf n},s)c({\bf n},s)+d^{\dagger}(-{\bf n},s)d(-{\bf n},s) 
\right) $$
$$ + \sum_{{\bf n},s}
\left\{-|{\bf p}_{\bf n}| s \cos{\theta_{\bf n}}+
\frac{2G}{L^3}{\mathcal{B}} s \sin{\theta_{\bf n}} \right\}
\left(c^{\dagger}({\bf n},s)d^{\dagger}(-{\bf n},s)+
d(-{\bf n},s)c({\bf n},s) \right) $$
$$ + {H'}_{int}  \eqno{ (2.9)} $$
where ${H'}_{int}$ is just the same as the one given in eq.(2.6). 
From this Hamiltonian, we get to know that the mass term is 
generated in the same way as the massive case. But we cannot 
make any renormalization since there is no mass term. Further, 
the new term is the only mass scale in this Hamiltonian since 
the coupling constant cannot serve as the mass scale. 
In fact, it is even worse than the dimensionless coupling constant case, 
since the coupling constant in the NJL model is proportional to the inverse 
square of the mass dimension. 
Thus, we define the new fermion mass $M_N$ by
$$ M_N = \frac{2G}{L^3}\mathcal{B}. \eqno{ (2.10)} $$
The Bogoliubov angle $\theta_{\bf n}$ can be determined from the following 
equation
$$ \tan \theta_{\bf n}  = {|{\bf p}_{\bf n}|\over M_N} . \eqno{ (2.11)} $$
In this case, the vacuum changes drastically since the original 
vacuum is trivial. 
 
Further, the constraints of eqs.(2.10) and (2.11) give rise to the equation 
that determines the relation between the induced fermion mass $M_N$ 
and the cutoff momentum $\Lambda$.  
$$ M_N={4G\over{(2\pi)^3}}\int^{\Lambda}d^3p {M_N\over{\sqrt{M_N^2+p^2}}} .
\eqno{(2.12)} $$
This equation has a nontrivial solution for $M_N$, and the vacuum 
energy becomes lower than the trivial vacuum ($M_N =0$). 
Therefore, $M_N$ can be expressed in terms of $\Lambda$ as
$$ M_N=\gamma \Lambda $$
where $\gamma$ is a simple numerical constant. 

It should be noted that the treatment up to now is exactly the same as 
the one given by Nambu and Jona-Lasinio \cite{q6}. Further, we stress 
that the induced fermion mass $M_N$ can never be set to zero, 
and it is always finite. 

\item{Boson mass}

The boson state $|B\rangle$  can be expressed as
$$ |B\rangle = \sum_{{\bf n},s}f_{\bf n}
c^{\dagger}({\bf n},s)d^{\dagger}(-{\bf n},s)|\Omega\rangle, 
 \eqno{(2.13)} $$
where $f_{\bf n}$ is a wave function in momentum space, and $|\Omega\rangle$ 
denotes the Bogoliubov vacuum state.
The equation for the boson mass $ {\cal M}$ for the NJL model  is written 
in terms of the Fock space expansion at the large $L$ limit 
$$ {\cal M}f(p) = 2E_{p}f(p)
-\frac{2G}{(2\pi)^3}\int^{\Lambda} d^3q  f(q)\left(1+
\frac{M^2}{E_{p}E_{q}}+\frac{{\bf p}\cdot {\bf q}}{E_{p}E_{q}}\right)  
 \eqno{(2.14)} $$
where  $M$ should be taken to be $M=m$ for the massive case, and 
$M=M_N$ for the massless case.  It is important to note that the fermion 
mass $M$ after the Bogoliubov transformation, therefore, cannot become zero. 

Here, again, we note that the RPA 
calculation gives the similar boson spectrum  
to the Fock space expansion. But we do not know whether the RPA calculation 
is better than the Fock space expansion or not, since the derivation 
of the RPA equation in field theory is not based on the fundamental 
principle. In principle, the RPA calculation may take into account the effect 
of the deformation of the vacuum in the presence of the particle and antiparticle. 
However, this is extremely difficult to do it properly, and indeed the RPA 
eigenvalue equation is not hermite, and thus it is not clear whether 
the effect is taken into account in a better way or worse. 
The examination and the validity of the RPA equation 
will be given else where. 

The solution of eq.(2.14) can be easily obtained, and the boson mass spectrum 
for the NJL model is shown 
in Fig. 1. Note that the boson mass is measured in units of the cutoff momentum $\Lambda$. 
As can be seen from the figure, there is a massive boson for some regions 
of the values of the coupling constant. Here, as we will see later, the NJL 
and the Thirring models are quite similar to each other. 
This is mainly because the current-current interaction is essentially a delta 
function potential in coordinate space. 
Indeed, as is well known, the delta function potential in one dimension can always bind 
the fermion and anti-fermion while the delta function potential in three dimension 
cannot normally bind them. Due to the finite cut off momentum, 
the delta function potential in three dimensions 
can make a weak bound state, depending on the strength of the coupling constant. 
This result of the delta function potential in quantum mechanics is 
almost the same as  what is just shown in Fig. 1. 

It should be noted that there is no serious difficulty of proving 
the non-existence of the massless boson. However, if it were to prove 
the existence of the massless boson, it would have been extremely difficult 
to do it. For the massless boson, there should be a continuum spectrum, and 
this continuum spectrum of the massless boson should be differentiated from 
the continuum spectrum arising from the many body nature of the system. 
This differentiation must have been an extremely difficult task. 
In fact, even if one finds a continuum spectrum which has, for example, 
the dispersion of $E=c_0 p^2$ as often discussed in solid state physics, 
one sees that the spectrum has 
nothing to do with the Goldstone boson.

\end{enumerate}

\vspace{1cm}

\item{\large Massless Thirring model}

Now, we discuss the Thirring model which is a two dimensional field theory with 
the current-current interaction \cite{q7}. 
For the massive Thirring model, it is discussed in detail in \cite{q9}, 
and therefore, we treat only the massless case in this paper. 

Since the massless Thirring model does not contain 
any scale parameter, it is necessary to introduce a scale or a cutoff 
momentum if one wants to discuss physical quantities. In this respect, 
for the massless Thirring model, any calculations with the regularization 
is physically meaningful. For example, the equivalence between the massless Thirring 
model and the massless boson system is quite well known and interesting, but 
this is mathematically important. However, physically, this equivalence is 
somewhat more complicated than expected 
since there is no scale introduced in the systems, and thus one cannot measure 
physical observables. 

The massless Thirring model is described by the following Lagrangian density
$$  {\cal L} = i \bar \psi  \gamma_{\mu} \partial^{\mu}  \psi 
  -{1\over 2} g j^{\mu} j_{\mu}   \eqno{ (3.1)} $$
where the fermion current $  j_{\mu} $  is given as 
$$  j_{\mu} = :\bar \psi  \gamma_{\mu} \psi : \eqno{ (3.2)}   $$
It is clear that this Lagrangian density has a chiral symmetry, and 
therefore, defining the chiral current by 
$$  j_{\mu}^5 = :\bar \psi \gamma_5 \gamma_{\mu} \psi : \eqno{ (3.3)}   $$
we see that the chiral current is conserved in the classical level 
$$  \partial^{\mu} j_{\mu}^5 =0  . \eqno{ (3.4)}   $$
In the recent paper, Faber and Ivanov \cite{q4} show that the true vacuum state 
has a chiral symmetry broken phase. They consider that the chiral symmetry 
breaking is spontaneous, and therefore they discuss the reason 
why the symmetry can be broken in the two dimensional field theory. 
Their discussions are concerned with the problem of Coleman's theorem.   

Here, we present the calculation which is based on the Bogoliubov 
transformation, and show that the chiral symmetry is indeed broken.  
However, we also show that there appears no massless boson in this 
regularized Thirring model.

Here, we first quantize the fermion field in a box $L$ 
$$ \psi (x) = {1\over{\sqrt{L}}} \sum_n  
\left(\matrix{a_n  \cr
     b_n \cr }\right)   e^{i{{2\pi nx}\over L}}   .  \eqno{ (3.5)} $$ 
In this case, the Hamiltonian of the massless Thirring model can be written as 
$$ H=\sum_{n}\left[p_{n}(a_{n}^{\dagger}a_{n}-b_{n}^{\dagger}b_{n})
+\frac{2g}{L}\tilde{j}_{1,p_{n}}\tilde{j}_{2,-p_{n}}\right], \eqno{ (3.6)} $$
where the fermion currents in the momentum representation $\tilde{j}_{1,p_{n}}$ and 
$\tilde{j}_{2,p_{n}}$ are given by
$$ \tilde{j}_{1,p_{n}}=\sum_{l}a_{l}^{\dagger}a_{l+n}  \eqno{ (3.7a)} $$
$$ \tilde{j}_{2,p_{n}}=\sum_{l}b_{l}^{\dagger}b_{l+n}. \eqno{ (3.7b)} $$
Now, we define new fermion operators by the Bogoliubov transformation, 
$$ c_n = e^{- {\cal A} } a_n e^{{\cal A}} = 
\cos \left({\theta_n\over 2}-{\pi\over 4}\right) a_n 
- \sin \left({\theta_n\over 2}-{\pi\over 4}\right) b_{n} \eqno{(3.8a)} $$
$$ d_{-n}^{\dagger} = e^{- {\cal A} } b_n e^{{\cal A}} 
= \cos \left({\theta_n\over 2}-{\pi\over 4}\right) b_n + 
\sin \left({\theta_n\over 2}-{\pi\over 4}\right) a_{n} \eqno{(3.8b)} $$
where the generator of the Bogoliubov transformation is given by
$$ {\cal A} = - \sum_n \left({\theta_n\over 2}-{\pi\over 4}\right)
 ( a^{\dagger}_n b_{n} -b^{\dagger}_{n}a_{n} )  . \eqno{(3.9)} $$
$ \theta_n$ denotes the Bogoliubov angle 
which can be determined by the condition that the vacuum energy is minimized. 
In this case, the new vacuum state is obtained as 
$$ \mid \Omega \rangle = e^{ - {\cal A} }  |0\rangle  .  \eqno{(3.10)}  $$

Now, we can obtain the new Hamiltonian under the Bogoliubov transformation, 
$$ H= \sum_{n}\Biggl[\left\{p_{n}\sin{\theta_{n}}
+\frac{g}{L}{\mathcal{B}}\cos{\theta_{n}}\right \}
(c_{n}^{\dagger}c_{n}+d_{-n}^{\dagger}d_{-n})\Biggl] $$
$$ + \sum_{ n}
\left\{-{ p}_{ n}\cos{\theta_{ n}}+
\frac{g}{L}{\mathcal{B} } \sin{\theta_{n}}   \right\} 
(c_{n}^{\dagger}d_{-n}^{\dagger}+d_{-n}c_{ n}) 
 +  H'  \eqno{ (3.11)}  $$

where $H'$ denotes the interaction terms of the Bogoliubov transformed 
state and is given in detail in \cite{q9}. Here, ${\mathcal{B}} $ is defined 
as ${\mathcal{B}}=\sum\cos \theta_{n}$, and the term  ${g\over L}{\mathcal{B}}$ 
corresponds to an induced mass. Therefore, we define the induced mass $M$ by
$$ M= \frac{g}{L}{\mathcal{B}}  . \eqno{ (3.12)} $$
The Bogoliubov angle $\theta_n$ can be determined by imposing the condition 
that the vacuum energy must be minimized. Therefore, we obtain
$$ \tan \theta_n  = {p_n\over M} . \eqno{ (3.13)} $$
In this case, we can express the self-consistency condition for $M$, and obtain  
$$ M= {g\over{\pi}}M \ln \left({\Lambda\over M} +
\sqrt{1+\left({\Lambda\over M}\right)^2 }
\right) \eqno{ (3.14)} $$
where $\Lambda$ denotes the cutoff. Since the massless Thirring model 
has no scale, we should measure all of the observables in terms of $\Lambda$. 
Therefore, we can express the induced mass $M$ in terms of $\Lambda$,
$$ M={\Lambda\over{\sinh \left({\pi\over g}\right)}} . \eqno{(3.15)} $$ 
Further, the vacuum energy $E_{vac}$ as measured from the trivial vacuum  is given 
$$ E_{vac}=-{L\over{2\pi}}{\Lambda^2\over{\sinh \left({\pi\over g}\right) }}
e^{-{\pi\over g}} . \eqno{(3.16)} $$
From this value of the vacuum energy, we get to know that the new vacuum energy 
is indeed lower than the trivial one. Therefore, the chiral symmetry is broken 
in the new vacuum state since the fermion becomes massive. 

Note that the present expression of the mass $M$ is somewhat different from the one 
given in Faber and Ivanov \cite{q4}. This is related to the fact that we take the cutoff 
momentum in the normal momentum while they used the cutoff in the Lorentz 
invariant fashion. 

In the same manner as \cite{q9}, we carry out the calculations of the spectrum 
of the bosons in the Fock space expansion. 

The boson state $|B\rangle$  can be expressed as
$$ |B\rangle = \sum_{n}f_{n}c_{n}^{\dagger}d_{-n}^{\dagger}|\Omega\rangle,
 \eqno{(3.17)} $$
where $f_{n}$ is a wave function in momentum space, and $|\Omega\rangle$ 
denotes the Bogoliubov vacuum state. The energy eigenvalue 
of the Hamiltonian for the large $L$ limit can be written as 
$$ {\cal M}f(p) = 2E_{p}f(p)
-\frac{g}{2\pi}\int dq f(q)\left(
1+\frac{M^2}{E_{p}E_{q}}+\frac{pq}{E_{p}E_{q}}\right) 
 \eqno{(3.18)} $$
where $ {\cal M}$ denotes the boson mass. 
$E_{p}$ is given as 
$$ E_{p} = \sqrt{M^2+p^2} . \eqno{(3.19)}  $$
Eq.(3.18) can be solved exactly as shown in \cite{q9}. First, we define $A$ and $B$ by
$$ A = \int_{-\Lambda}^{\Lambda} dp f(p)  \eqno{(3.20a)}  $$ 
$$ B = \int_{-\Lambda}^{\Lambda} dp \frac{f(p)}{E_{p}}. \eqno{(3.20b)} $$
Using $A$ and $B$, we can solve Eq. (3.18) for $f(p)$ and obtain
$$ f(p) = \frac{g/2\pi}{2E_{p}-\mathcal{M}}\left(A+\frac{m^2}{E_{p}}B\right).
 \eqno{(3.21)} $$
Putting this $f(p)$ back into Eqs. (3.20), we obtain the matrix equations
$$ A = \frac{g}{2\pi}\int_{0}^{\Lambda} \frac{2dp}{2E_{p}-\mathcal{M}}
\left(A+\frac{m^2}{E_{p}}B\right) \eqno{(3.22a)}  $$
$$ B = \frac{g}{2\pi}\int_{0}^{\Lambda} \frac{2dp}{(2E_{p}-{\mathcal{M}})E_{p}}
\left(A+\frac{m^2}{E_{p}}B\right). \eqno{(3.22b)} $$
Since the model is already regularized, we can easily calculate the boson 
spectrum which is given in Fig. 2 as the function of the coupling 
constant $g/\pi$. 
As can be seen from Fig. 2, there is no massless boson 
in this spectrum even though the boson mass for the very small coupling constant $g$ 
is exponentially small. Here, we also carry out the  RPA calculation for the massive 
Thirring model. 
However, the spectrum predicted by the RPA calculation 
somewhat deviates from the other calculated results \cite{q12,q13,q14,q15} 
in the massive Thirring model. Since the derivation of the RPA equation 
in field theory is not based on the fundamental principle, we do not know 
to what extent the calculation of the RPA is reliable and 
the validity of the RPA equation will be 
examined else where. 

Now, we discuss the chiral current conservation. In order to examine it, 
we evaluate the following equation, 
$$ i{\dot Q}_5 =[H, Q_5]  \eqno{(3.23)} $$
where $Q_5$ is defined as 
$$ Q_5(t)=\int j_0^5 (x,t) dx 
=\int \bar \psi  \gamma_0\gamma_5 \psi dx . \eqno{(3.24)} $$
Clearly, we can show that the right hand side of eq.(3.23) vanishes 
for the massless Thirring Hamiltonian of eq.(3.4). However, since 
the Bogoliubov transformation is unitary, $[H, Q_5]$ is invariant 
and thus it remains to be zero. Therefore, the chiral current 
is still  conserved after the Bogoliubov transformation. 

Here, we also calculate the fermion condensate $C$ for this vacuum state \cite{q11}, 
and obtain 
$$ |C|= \langle \Omega \mid {1\over L} \int \bar \psi \psi dx \mid \Omega \rangle 
      = {M\over{g}}      \eqno{(3.25)} $$
which is a finite value. 
Therefore, the massless Thirring model has a chiral symmetry broken 
vacuum, but, the chiral current is conserved. 
Therefore, this situation contradicts the Goldstone theorem 
since there appears no massless boson in this symmetry breaking.  
This point will be discussed in detail in the next section.

\vspace{1cm}

\item{\large Goldstone Theorem}

Here, we discuss the Goldstone theorem in connection with the massless 
NJL and the Thirring  models. In the formal proof 
of the Goldstone theorem \cite{q1,q2}, one assumes 
the existence of the vacuum expectation value of the following commutation relation, 
$$ \langle \Omega \mid [Q_5(t), \phi (0)] \mid \Omega \rangle  \neq 0  
\eqno{(4.1)} $$
where the boson field $\phi (x)$ must be constructed by the fermion fields. 
Now, inserting intermediate boson states, one obtains 
$$ \sum_n \delta ({\bf p}_n) \left[ e^{iE_nt} \langle \Omega \mid j^5_0(0) 
\mid n \rangle  \langle n \mid \phi(0) \mid \Omega \rangle -
e^{-iE_nt} \langle \Omega \mid \phi(0) 
\mid n \rangle  \langle n \mid j^5_0(0)  \mid \Omega \rangle \right]  \neq 0  
\eqno{(4.2)} $$
The right hand side of eq.(4.2) is non-vanishing, and time-independent. 
Thus, eq.(4.2) can be satisfied only if there exists an intermediate state 
with $E_n=0$ for ${\bf p}_n=0$. This is the proof of the Goldstone theorem, and 
it is indeed valid for boson field theory. 
One can also obtain the same result from 
rewriting the Lagrangian density at the new vacuum point for boson field theory models. 
This is reasonable since the Goldstone boson is understood 
as a kinematical effect. 

On the other hand, the fermion field theory models  are quite different, and 
one cannot obtain the Goldstone boson even if one rewrites the Lagrangian 
density at the new vacuum point. This is due to the fact that 
the Goldstone boson must be constructed by the fermion and anti-fermion fields, 
and therefore it should involve the dynamics. 
But it has been believed that the proof 
based on eqs.(4.1) and (4.2) holds for the fermion field theory models  as well. 
However, one must be careful for the fermion field theory models  with 
the regularization. In the above proof, the use of 
the translational property of a boson plays an essential role and it is assumed 
for the boson field $\phi (x)$ and 
 the chiral current $j_{\mu}^5(x)$ even if they are constructed by fermions. 
However,  the boson field $\phi(x)$ together with the current $j_{\mu}^5(x)$ 
must be regularized by the point splitting, 
$$  j_{\mu}^5(x) = \bar \psi(x) \gamma_5 \gamma_{\mu} 
\psi(x+\epsilon)  \eqno{ (4.3a)}   $$
$$  \phi(x) = \bar \psi(x) \gamma_5  
\psi(x+\epsilon)  \eqno{ (4.3b)}   $$
where we have to keep the $\epsilon$ finite. In the NJL as well as 
the Thirring  models, the $\epsilon$ should be related to the cutoff momentum $\Lambda$ 
in some way or the other. 
In this case, one cannot 
make use of the translation property of $\phi(x)$ 
$$  \phi (x) = e^{ipx}  \phi(0) e^{-ipx}  .  $$
Instead, $  \phi (x)$ is only written as 
$$  \phi (x) = e^{ipx} \bar \psi(0) \gamma_5  e^{ip\epsilon} 
\psi(0) e^{-ip(x+\epsilon)} , \eqno{ (4.4)}   $$
and  one cannot obtain the same equation as eq.(4.2) for fermion 
fields with the regularization. Therefore one cannot claim the existence 
of the massless boson any more. 
In the massless NJL as well as the Thirring  models, 
one needs to regularize the fermion 
field, and  therefore, one cannot claim that the Goldstone theorem should hold true. 
In fact, as we saw in the previous sections, there appears no massless boson, 
and this is just what is found here.  

Further, we should note that no fermion field theory model except the NJL 
model predicts a Goldstone boson. It is often discussed that pion may 
well be a Goldstone-like boson. But for this, one should be careful. 
In four dimensional QCD, the coupling constant has no dimension, and 
therefore, all the physical mass must be measured by the quark mass 
if the quarks are massive. In this case, there is no chiral symmetry, 
and there is no need to discuss the Goldstone boson. It is believed 
that the pion mass is too light compared with other mesons. 
However, it is quite large if one measures it in terms 
of the quark mass which should be around 10 MeV. In this 
respect, the pion mass should be considered as an 
object which has nothing to do with the chiral symmetry breaking. 
If the quarks were massless, then the chiral symmetry should be 
broken in the new vacuum, and any physical observables like meson masses 
should be measured by the cut off momentum $\Lambda_{QCD}$. But 
this story has nothing to do with nature in real QCD.

\vspace{1cm}

\item{\large Conclusions}

We have presented the chiral symmetry breaking of the massless NJL 
and the Thirring  models. 
The new vacuum has a finite chiral condensate. Also, 
it is  shown that the chiral symmetry breaking here does not 
accompany a massless boson. This is not consistent with the Goldstone 
theorem since the chiral current 
conservation still holds in the Bogoliubov transformed vacuum with the regularization. 

This inconsistency is resolved in that the Goldstone theorem turns 
out to be invalid for the fermion field theory models  with the regularization. 
This is simply due to the fact that the proof of the Goldstone theorem 
is essentially based on the use of the translational property of the boson, 
which, however, cannot be valid for the fermion and antifermion bound state 
due to the point splitting regularization.

To summarize the chiral symmetry breaking in fermion field theory models 
of the NJL and the Thirring, 
we state that the new vacuum indeed violates the chiral symmetry, and 
it has a finite condensate value. The fermion acquires the finite mass, and 
the rest of the theory becomes just the massive fermion field theory 
with the same interactions as the massive fermion case. 
This completes the symmetry breaking business. 
There is no space and no freedom left for the Goldstone boson. The  
massive fermion field theory models of the NJL and the Thirring  
give a massive boson. 
The Thirring model has an exponentially small boson mass 
for the small coupling constant region while the boson mass of the NJL model 
strongly depends on the strength of the coupling constant. 
Indeed, below the critical value of the coupling constant, there is 
no bosonic bound state. 
However, the boson mass has little to do with the symmetry breaking phenomena, 
but it is determined by the coupling constant and the cutoff $\Lambda$. 

What is then the main difference between the boson field theory and the fermion 
field theory ? In boson field theory models, one assumes 
that the potential term $U(|\phi|)$ 
for the boson field has a nontrivial minimum, which has infinite degenerate states. 
This degeneracy of the "potential vacuum" is resolved when one considers 
the kinetic energy terms of the boson field. At this point, the symmetry is broken 
and one obtains the new vacuum state with a massless boson. Here, one notices 
that, if there were no nontrivial minimum in the potential vacuum, then there 
should exist no Goldstone boson. Further, if one finds a system whose $total$ 
$Hamiltonian$ has the vacuum with infinite degenerate states, 
then there is no way to resolve the degeneracy, 
and the vacuum should stay as it is. The fermion field theory models which we treat 
in this paper have no nontrivial minimum of the potential vacuum, and the vacuum 
is found only when one considers the total Hamiltonian of the system. Therefore, 
there is no place one finds any massless boson degree of freedom for the fermion 
field theory models. 

\vspace{0.5cm}

We would like to thank  T. Nihei and K. Yazaki for helpful 
discussions and T. Asaga for critical reading of the manuscript.

\newpage
\begin{figure}[h]
\hskip 3cm
\epsfig{file=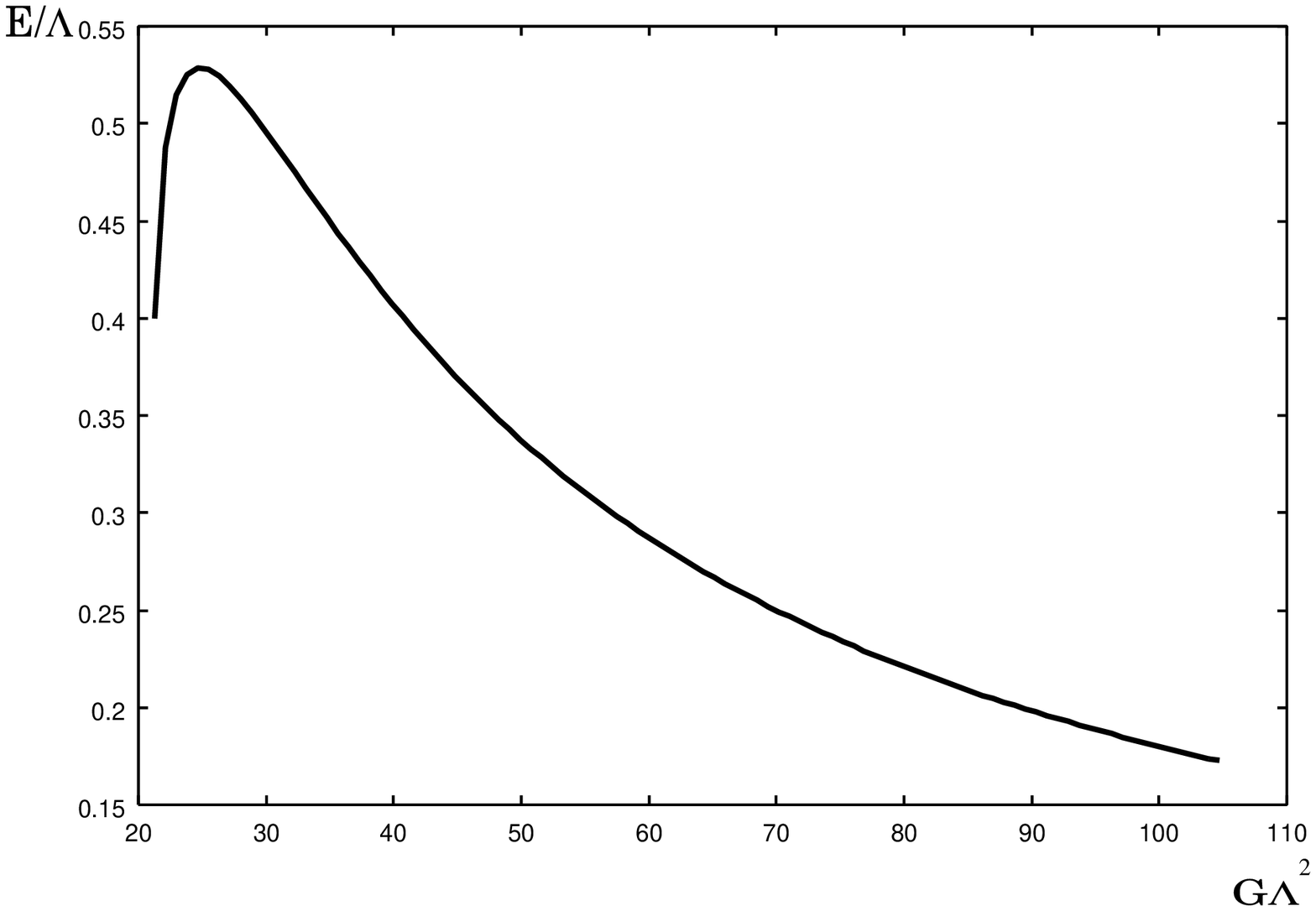, width=14cm}
\end{figure}

\vspace{1cm}

Fig. 1: The boson mass for the NJL model is plotted as the function of $G\Lambda^2$. \\
\qquad \quad \  It is measured by the cutoff $\Lambda$. 

\newpage

\begin{figure}[h]
\hskip 3cm
\epsfig{file=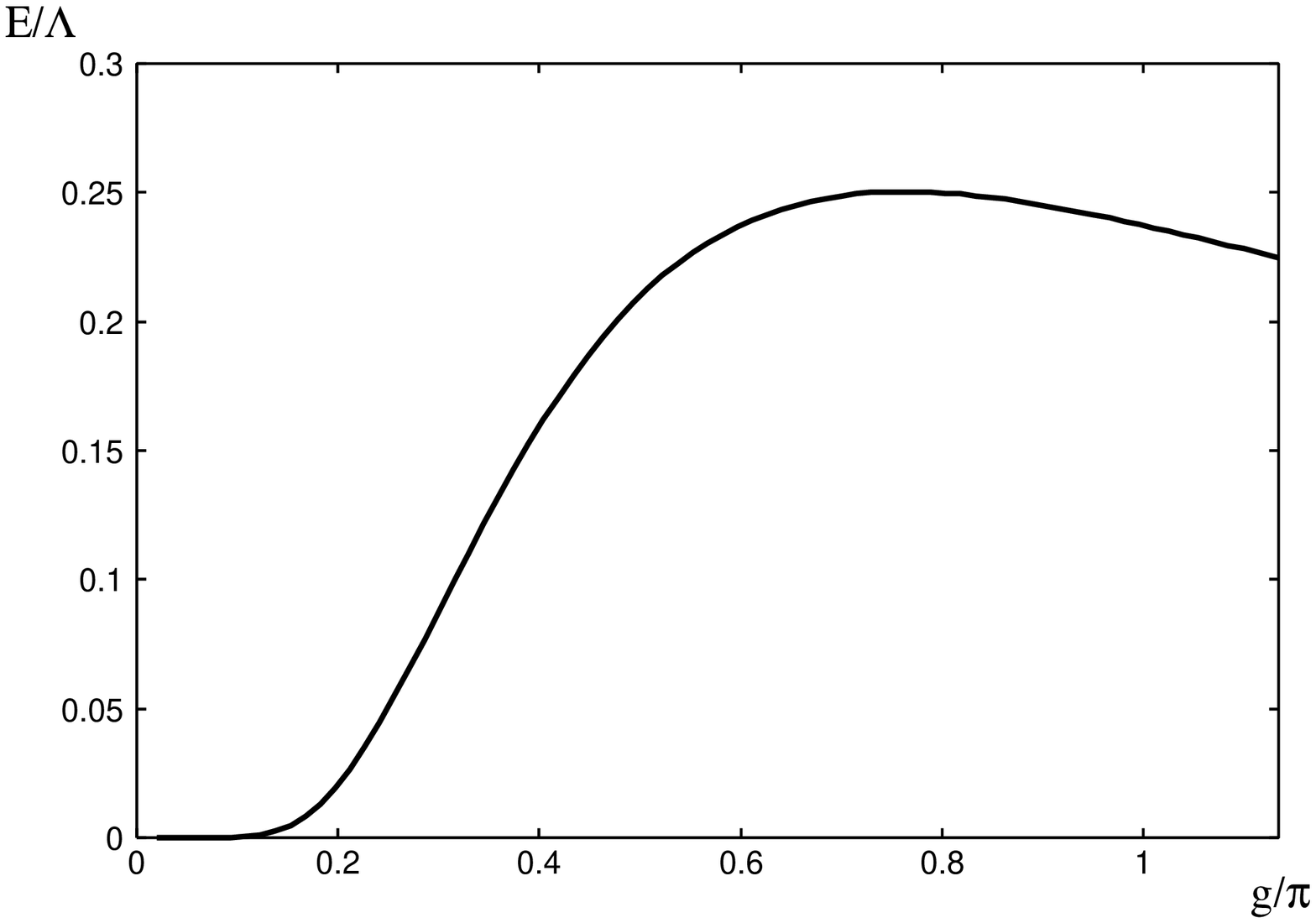, width=14cm}
\end{figure}

\vspace{1cm}

Fig. 2: The boson mass for the massless Thirring model is plotted \\
\qquad \quad \  as the function of $g/\pi$. 
It is measured by the cutoff $\Lambda$.

\end{enumerate}

\newpage

\begin{thebibliography}{99}

\bibitem{q1}
J. Goldstone, Nuovo Cimento, {\bf 19}, 154 (1961)

\bibitem{q2}
 J. Goldstone, A. Salam and S. Weinberg, Phys. Rev. {\bf 127}, 965 (1962) 


\bibitem{q6}
Y. Nambu and G. Jona-Lasinio, Phys. Rev. {\bf 122}, 345 (1961) 



\bibitem{q07}
S. Klevansky, Rev. Mod. Phys. {\bf 64}, 345 (1992), 649 


\bibitem{q7}
W. Thirring, Ann. Phys. (N.Y) {\bf 3}, 91 (1958) 


\bibitem{q9}
M. Hiramoto and T. Fujita,  Phys. Rev. {\bf D66}, 045007 (2002)


\bibitem{q3}
S. Coleman, Comm. Math. Phys. {\bf 31} (1973), 259

\bibitem{q34}
T. Fujita, M. Hiramoto and T. Homma, \\
"New spectrum and condensate in two dimensional QCD", submitted to Ann. Phys. 



\bibitem{q4}
M. Faber and A.N. Ivanov, Eur. Phys. J. {\bf C20}, 723 (2001)



\bibitem{q11}
T. Tomachi and T. Fujita,  Ann. Phys. {\bf 223}, 197 (1993)


\bibitem{q12}
T. Fujita and A. Ogura, Prog. Theor. Phys. {\bf 89},  23 (1993)


\bibitem{q13}
T. Fujita, Y. Sekiguchi and K. Yamamoto, Ann. Phys. {\bf 255},  204 (1997)


\bibitem{q14}
T. Fujita and M. Hiramoto, Phys. Rev. {\bf D58}, 125019 (1998)


\bibitem{q15}
R. F. Dashen, B. Hasslacher and A. Neveu, Phys. Rev. {\bf D11}, 3432 (1975) 

\end{thebibliography}
\end{document}